\documentclass[usenatbib]{mnras}

\usepackage{graphicx}
\usepackage{amsmath}
\usepackage{amssymb}
\usepackage{booktabs}
\usepackage{array}
\usepackage{hyperref}

\def\ltsima{$\; \buildrel < \over \sim \;$}
\def\gtsima{$\; \buildrel > \over \sim \;$}
\def\simlt{\lower.5ex\hbox{\ltsima}}
\def\simgt{\lower.5ex\hbox{\gtsima}}
\def\p2Y{\;_2Y}
\def\m2Y{\;_{-2}Y}

\def\Hunit{~{\rm km} \ {\rm s}^{-1}{\rm Mpc}^{-1}}

\def\mk2{\mu {\rm K}^2}
\def\Planck{\it Planck\rm}

\def\LCDM{$\Lambda$CDM}

\def\rmd{\mathrm{d}}
\def\zeff{z_{\rm eff}}
\def\rfid{r_{d, {\rm fid}}}

\def\pmb#1{\setbox0=\hbox{#1}%
     \kern-.025em\copy0\kern-\wd0
     \kern.05em\copy0\kern-\wd0
     \kern-.025em\raise.0433em\box0}

\def\beglet{
  \addtocounter{equation}{1}%
  \setcounter{parentequation}{\value{equation}}%
  \setcounter{equation}{0}%
  \def\theequation{\arabic{parentequation}\alph{equation}}%
  \ignorespaces
}
\def\endlet{
  \setcounter{equation}{\value{parentequation}}%
  \def\theequation{\arabic{equation}}%
}

\title[Model independent $H(z)$ reconstruction using the cosmic inverse distance ladder]
{Model independent $H(z)$ reconstruction using the cosmic inverse distance ladder}

\author[Pablo Lemos, Elizabeth Lee, George Efstathiou, Steven Gratton]{Pablo Lemos\thanks{E-mail: pl411@cam.ac.uk}, Elizabeth Lee, George Efstathiou, Steven Gratton \\
 Kavli Institute for Cosmology Cambridge and 
Institute of Astronomy, Madingley Road, Cambridge, CB3 OHA.}

\date{Accepted XXX. Received YYY; in original form ZZZ}
\pubyear{2018}

\begin{document}
\label{firstpage}
\pagerange{\pageref{firstpage}--\pageref{lastpage}}
\maketitle

\begin{abstract} 

Recent distance ladder determinations of the Hubble constant $H_0$
disagree at about the $3.5\sigma$ level with the value determined from
\Planck\ measurements of the cosmic microwave background (CMB)
assuming a $\Lambda$CDM\textsuperscript{\footnotemark} cosmology. This discrepancy has
prompted speculation that new physics might be required beyond that
assumed in the \LCDM\ model.  In this paper, we apply the inverse
distance ladder to fit a parametric form of $H(z)$ to baryon acoustic
oscillation (BAO) and Type Ia supernova (SNe) data together with priors on
the sound horizon at the end of the radiation drag epoch, $r_d$. We
apply priors on $r_d$, based on inferences from either \Planck\ or the Wilkinson
Microwave Anistropy Probe (WMAP), and demonstrate that these values are
consistent with CMB-independent determination of $r_d$ derived from
measurements of the primordial deuterium abundance, BAO and supernova
data assuming the \LCDM\ cosmology.  The $H(z)$ constraints that we
derive are independent of detailed physics within the dark sector at
low redshifts, relying only on the validity of the Friedmann-Robertson-Walker (FRW) metric of
General Relativity. For
each assumed prior on $r_d$, we find consistency with the inferred
value of $H_0$ and the \Planck\ \LCDM\ value and corresponding tension
with the distance ladder estimate.

\vskip 0.3 truein

\end{abstract}

\begin{keywords} % commenting results in missing thanks notes!
cosmology: observations, distance scale, cosmological parameters, large-scale structure of Universe
\end{keywords}

\footnotetext{Here \LCDM\ refers to a spatially flat FRW cosmology dominated by cold dark matter and a cosmological
constant at the present date with Gaussian initial adiabatic fluctuations characterised by a power law  spectrum.}

\section{Introduction}
\label{sec:intro}

The \Planck\ satellite has provided strong evidence in support of the
\LCDM\ cosmology and has measured the six parameters that define this
model to high precision \citep[][hereafter P14 and P16 respectively]{Planckparams14,
  Planckparams16}. In particular, P16\footnote{This value is for
  the full temperature and polarization analyis in P16.
 It is consistent with the value $H_0= 67.36 \pm 0.54 \Hunit$
  from the latest \Planck\ analysis \citep{Planck:2018} derived for the
  TT,TE,EE+lowE+lensing likelihood combination.} found a value of the Hubble
constant of $H_0= 67.27 \pm 0.66 \Hunit$. As pointed out in P16, other data combinations give
similar values of $H_0$, for example combining WMAP and BAO data gives
$H_0= 68.0 \pm 0.7 \Hunit$.  A `low' value of $H_0$ is therefore not
solely driven by high multipole CMB anistropies measured by
\Planck\ but is necessary if the \LCDM\ cosmology is to fit a range of
cosmological data.
% \citep[see also][]{Addison:2018}.

In contrast, direct measurements of the cosmic distance scale have
consistently found a higher value of $H_0$. The
SH0ES\footnote{Supernovae and $H_0$ for the Equation of State.}
project uses Cepheid period-luminosity relations, together with local
distance anchors, to calibrate distances to Type Ia SNe host
galaxies. The SH0ES programme has reported measurements of $H_0$ of
increasing precision over the last few years \citep{Riess:2009,
  Riess:2011, Riess:2016, Riess:2018}. The latest value from the SH0ES
collaboration\footnote{As this work was nearing completion,
  \cite{Riess:2018c} reported new Hubble Space Telescope photometry of
  long period Milky Way Cepheids. Together with GAIA parallaxes
  \citep{Gaia:2018} these measurements increase the tension between
  \Planck\ and the distance ladder estimate of $H_0$ to $3.8\sigma$.}
is $H_0 = 73.48 \pm 1.66 \Hunit$ \citep[][hereafter R18]{Riess:2018}, which is consistent with but has a much smaller error than
earlier determinations from the Hubble Space Telescope key project \citep{Freedman:2001}.

The $3.5\sigma$ difference between the SH0ES determination of
$H_0$ and the value inferred from \Planck\ for the \LCDM\ cosmology is
one of the most intriguing problems in modern cosmology. Perhaps
unsurprisingly, there have been many attempts to solve the problem by
introducing new (and sometimes highly speculative) physics
\citep[e.g.][]{Wyman:2014, Zhao:2017, DiValentino:2018a, Sola:2017,
  DiValentino:2018b}. There have also been several reanalyses of the
SH0ES data \citep{Efstathiou:2014, Cardona:2017, Zhang:2017,
  Follin:2018} which, apart from minor details, agree well with the
analyses by the SH0ES collaboration, though \cite{Feeney:2018a}
conclude that the Gaussian likelihood assumption used in the SH0ES
analysis may overestimate the statistical significance of the
discrepancy.

In this paper, we apply the inverse ladder
\citep{Percival:2010, Heavens:2014, Aubourg:2015, Cuesta:2015, Bernal:2016, Abbott:2017, Verde:2017} to derive
an estimate of $H_0$.  In our application, we combine SNe data
from the Pantheon sample \citep{Scolnic:2017} with BAO measurements
from the 6dF Galaxy Survey (6dFGS) \citep{Beutler:2011}, Baryon
Oscillation Spectroscopic Survey (BOSS) \citep{Alam:2016} and Sloan
Digital Sky Survey (SDSS) quasars \citep{Bautista:2017, Bourboux:2017,
  Zarrouk:2018}. To calibrate the inverse distance ladder, we impose
priors on the sound horizon at the end of the radiation drag epoch,
$r_d$. However, instead of assuming a particular cosmological model,
we fit a flexible parametric  model describing the evolution of the Hubble
parameter  $H(z)$. The FRW metric of General Relativity then fixes
the luminosity distance $D_L(z)$ in terms of $H(z)$; the extrapolation
of $H(z)$ to $z=0$ is then independent of the low redshift properties
of dark matter and dark energy,  as in the important analysis of \cite{Heavens:2014}. 

The analysis presented here is similar
to recent analyses by \cite{Feeney:2018}, who parameterized $D_L(z)$
with a third-order Taylor expansion (characterized by the deceleration
and jerk parameters $q_0$ and $j_0$), by \cite{Joudaki:2017b}, who
parameterized $H(z)$ on a discrete grid in $z$ and by \cite{Bernal:2016}
 who reconstruct $H(z)$ by interpolating piece-wise cubic splines specified
by a small number of knots.  In this paper, we
parameterize $H(z)$ as a smooth function of redshift. Our analysis is
closely related to that of \cite{Bernal:2016}, except that we use more recent
(and more constraining) BAO and supernova data to extrapolate to a value of
$H_0$ rather than fixing the sound horizon, and we  demonstrate explicitly that the discrepancy with the direct measurement of $H_0$ is insensitive
to whether the BAO scale is normalized using priors on the sound horizon 
derived from  \Planck\ or WMAP.

The layout of our paper is as follows: In Section \ref{sec:idl} we
introduce our parameterization of $H(z)$ and the priors on $r_d$ that
we use to calibrate the distance scale. The datasets used in this
analysis are described in Section \ref{sec:data} and our results are
presented in Section \ref{sec:results}. Section \ref{sec:conc}
presents our conclusions.

\section{Inverse distance ladder}
\label{sec:idl}

\subsection{$H(z)$ parameterizations}

According to General Relativity, the Hubble parameter $H(z)$ fixes the
luminosity distance $D_L(z)$ and comoving angular diameter distance
$D_M(z)$ according to
\begin{equation}
    D_L(z) = c(1+z)\int_0^{z}{dz^\prime \over H(z^\prime)}, \\ \quad  D_M(z) = {D_L(z) \over (1+z)},  \label{idl1}
\end{equation}
where we have assumed that a spatially flat geometry is an accurate description of our Universe.
We adopt the following parameteric form for $H(z)$:
\begin{equation}
H^2(z) = H_{\rm fid}^2 \left[A(1+z)^3 + B + C z + D (1+z)^{\epsilon} \right],   \label{idl2}
\end{equation}
with $A, B, C, D$ and $\epsilon$ as free parameters. We refer to this parameterization as the `epsilon' model. The normalising
factor $H_{\rm fid}$ is fixed at $H_{\rm fid} = 67 \Hunit$ and is introduced so that the free parameters $A$ to $D$ are dimensionless and of order unity. 
In the base \LCDM\ cosmology, 
\begin{equation}
H(z) = H_0\left[ \Omega_m (1+z)^3 + (1-\Omega_m) \right]^{1/2}, \label{idl3}
\end{equation}
where $\Omega_m$ is the present day total matter density in units of the critical density. Equation (\ref{idl3})
applies at low redshifts when contributions to the energy density from photons and neutrinos can be ignored. This equation is
reproduced by the parametrization of equation (\ref{idl1}) if
\begin{eqnarray}
&A& = \left( {H_0 \over H_{\rm fid}} \right)^2 \Omega_m, \ \  B = \left ( {H_0 \over H_{\rm fid}} \right)^2 (1 - \Omega_m), \nonumber \\
&C&=D=0,  \epsilon \ne 0, \label{idl4}
\end{eqnarray}
with a degeneracy between $B$ and $D$ for $\epsilon=0$.

The base \LCDM\ model assumes that dark energy is a cosmological constant with equation of state $w = p/\rho = -1$.
In models of evolving dark energy, the equation of state is often parameterized as
\begin{equation}
w(z) = w_0 + w_a \frac{z}{1+z}.  \label {idl5}
\end{equation}
With this equation of state, and arbitrary curvature $\Omega_k$, 
\begin{eqnarray}
\label{elcdm}
 H^2(z)\hspace{-2mm} &=& \hspace{-2mm}  H^2_0 [  \Omega_m(1+z)^3 + \Omega_k (1+z)^2 \qquad \qquad  \nonumber \\
 & &  \quad  + \ \Omega_{\rm DE}(1+z)^{-3(1+w_0+w_a)} e^{-3 w_a z/(1+z)}],  \qquad 
\end{eqnarray}
where $\Omega_{\rm DE} = 1-\Omega_{m} - \Omega_{k}$. In our  application of the inverse distance ladder, the data that we
use spans the redshift range $0.1-2.4$ (see Section \ref{sec:data}). Over this redshift range, the parameterization
of equation (\ref{idl2}) accurately reproduces equation (\ref{elcdm}) for extreme values of $w_0$, $w_a$ and $\Omega_{k}$.
Provided $H(z)$ is a smoothly varying function of $z$, with no abrupt jumps,  the epsilon model provides an accurate
description of the evolution of $H(z)$ in a wide variety of theories involving dynamical dark energy and interactions
between dark energy, dark matter and baryons.

As we will see in Section \ref{sec:results}, the parameters of the epsilon model are strongly degenerate. We have therefore also implemented a simpler parameterization, which we refer to as the `log' model:
\begin{equation}
\label{log}
H^2(z) = H_{\rm fid}^2 \left[A^\prime(1+z)^3 + B^\prime + C^\prime z + D^\prime {\rm ln}(1+z)\right].
\end{equation}
This is a less flexible parameterisation than the epsilon model but the four free parameters in equation (\ref{log})
are less degenerate. In fact, we will find that the data constrain $H(z)$ to be so close to the form expected in the base
\LCDM\ cosmology that the epsilon and log models give nearly identical results for $H_0$.

\subsection{The sound horizon}

\begin{table*}
\centering
\begin{tabular}{c c  l l} 
 \hline
 Dataset & $\zeff$ &  Measurement & Constraint\\
 \hline 
 6dFGS & 0.106   &$r_d/D_V(\zeff)$ & $0.336 \pm 0.015$ \\ 
 BOSS DR12 & 0.38  & $D_M(\zeff) \rfid/r_d$ & $1512 \pm 25$ \  Mpc\\ 
  &  &   $H(\zeff) r_d/\rfid$  & $81.2 \pm 2.4$  $\Hunit$ \\ 
  & 0.51 &  $D_M(\zeff) \rfid/r_d $ & $1975 \pm 30$ \   Mpc\\ 
  &  &   $H(\zeff) r_d/\rfid$ & $90.9 \pm 2.3$  $\Hunit$ \\
    & 0.61  & $D_M(\zeff) \rfid/r_d $ & $2307 \pm 37$  \ Mpc\\ 
  &  &   $H(\zeff) r_d/\rfid$ & $99.0 \pm 2.5$  $\Hunit$ \\
 eBOSS DR14 QSO & 1.52 &$D_A(\zeff) \rfid/r_d$ & $1850^{+90}_{-115}$ \  \ Mpc \\ 
  &    &$H(\zeff) r_d/\rfid$ & $159^{+12}_{-13}$ \ \ \ \  $\Hunit$  \\ 
 BOSS DR12 ${\rm Ly\alpha}$ & 2.33 & $D_M(\zeff) /r_d $ & $37.8 \pm 2.1$ \\ 
  &  &   $c / (H(\zeff) r_d)$ & $9.07 \pm 0.31$ \\ 
BOSS DR12 QSOx${\rm Ly\alpha}$ & 2.40 & $D_M(\zeff)/r_d$ & $35.7 \pm 1.7$ \\ 
  &  &  $c / (H(\zeff) r_d)$ & $9.01 \pm 0.36$  \\
  \hline
\end{tabular}
\caption{BAO measurements used in this paper. $\zeff$ gives the effective redshift for each measurement.
The BOSS DR12 and eBOSS DR14 QSO analyses adopt a fiducial sound horizon of $\rfid = 147.78 \ {\rm Mpc}$. 
 Note that \protect\cite{Beutler:2011} use the \protect\cite{Eisenstein:1998}  formulae to calculate $r_d$.}
\label{table:BAO}
\end{table*}

The principal datasets used in this analysis are Type Ia supernovae, for which we require the luminosity distance $D_L(z)$,
and BAO data which return joint estimates of $D_M(z)/r_d$ and $H(z)r_d/c$. Here $r_d$ is the sound horizon at the epoch $z_d$
when baryons decouple from the photons:
\begin{equation}
\label{rd}
r_d = \int_{z_d}^{\infty} \frac{c_s(z)}{H(z)} \rmd z, 
\end{equation}
where $c_s$ is the sound speed in the photon-baryon fluid, given by
\begin{equation}
\label{cs}
c_s^2(z) = \frac{c^2}{3} \left[1+\frac{3}{4}\frac{\rho_b(z)}{\rho_{\gamma}(z)} \right]^{-1},
\end{equation}
where $\rho_b$ and $\rho_{\gamma}$ are the energy densities of baryons and radiation respectively. CMB experiments such as
\Planck\ and WMAP \citep{WMAP:2013} lead to precise determinations of $r_d$. From the 2015 Planck Legacy Archive (PLA)  tables\footnote{\url{http://www.cosmos.esa.int/web/planck/pla}.} we have
\beglet
\begin{eqnarray}
r_d &=& 147.27 \pm 0.31 \ {\rm Mpc}, \quad  {\rm Planck}, \label{rd1a}\\
r_d &=& 148.5 \pm 1.2 \ {\rm Mpc},  \quad  \ \ \   {\rm WMAP9}, \label{rd1b}
\end{eqnarray}
\endlet where the \Planck\ value is for the likelihood combination
TE+TE+EE+lowTEB in the notation of P16.\footnote{This is consistent with the 
value $r_d = 147.09 \pm 0.26 \ {\rm Mpc}$ derived for the 
  TT,TE,EE+lowE+lensing likelihood combination \citep{Planck:2018}.} We use the PLA value for the
nine-year WMAP estimate, rather than the value quoted in
\cite{WMAP:2013}, since (\ref{rd1b}) is calculated consistently using
the Boltzmann solver {\tt CAMB} \citep{Lewis:2000}.

The estimates of $r_d$ in (\ref{rd1a}) and (\ref{rd1b}) are extremely
insensitive to physics at low redshifts \citep{Cuesta:2015} (since the
physical densities $\Omega_mh^2$ and $\Omega_ch^2$ which enter in
equation (\ref{rd}) are fixed mainly by the relative heights of the
CMB acoustic peaks) but assume the base \LCDM\ cosmology at high
redshifts. By using these values as priors in the inverse distance
ladder, we are implicitly assuming that the base \LCDM\ model is
correct at high redshift though we allow deviations from the model
at low redshifts via the parameterizations of equations (\ref{idl2})
or (\ref{log}). However, as discussed in P14 and P16, the parameters
of the base \LCDM\ found by \Planck\ are consistent with Big Bang
Nucleosynthesis (BBN) constraints on $\Omega_bh^2$ inferred from deuterium
abundance measurements in low metallicity systems at high redshift
\citep{Cooke:2014, Cooke:2016, Cooke:2018}. As emphasised by
\cite{Addison:2018}, BBN constraints can be used together with BAO
data to provide a consistency check of $r_d$ and $H_0$ assuming the
base \LCDM\ model. We will revisit this constraint in Section
\ref{subsec:rd}.

\section{Data}
\label{sec:data}

The BAO measurements used in this paper are summarized in Table \ref{table:BAO}. We use the BAO measurements from the 6dF Galaxy Survey (6dFGS) \citep{Beutler:2011} which constrains $r_d/D_V$, where 
\begin{equation}
D_V(z) = \left [ D^2_M(z) {cz \over H(z)} \right ]^{1/3}.
\end{equation}
Note that \citep{Beutler:2011} use the \cite{Eisenstein:1998} formulae
to calculate $r_d$. The {\tt CAMB} code gives values that are lower by
a factor $1.027$ and so the \cite{Beutler:2011} numbers in Table 1
have been corrected to account for this difference (for a more
detailed discussion see Appendix B of \cite{Hamann:2010}).  We use the
BOSS DR12 consensus BAO measurements \citep{Alam:2016} on $D_M(z)$ and
$H(z)$ in three redshift bands together with the associated $6\times6$
covariance matrix\footnote{ BAO\_consensus\_covtot\_dM\_Hz.txt
  downloaded from
  \url{http://www.sdss3.org/science/BOSS\_publications.php}.}. We also
use the eBOSS BAO measurements from quasars in DR14
\citep{Zarrouk:2018}, from BOSS DR12 analyses of Lyman-$\alpha$
absorption in quasar spectra \citep{Bautista:2017} and BAO constraints
from a ${\rm Ly\alpha}$-quasar cross-correlation analysis with BOSS
DR12 \citep{Bourboux:2017}. The high redshift measurements are less
accurate than the BOSS DR12 galaxy measurements, but serve to anchor
the parametrizations (\ref{idl2}) and (\ref{log}) at redshifts greater
than unity. Note also that since the likelihoods for these high redshift measurements  were not available to us, and these data are relatively unimportant
for fixing $H_0$,  we sampled over
$D_M(z)$ and $H(z)$ assuming that they are Gaussian distributed and
uncorrelated.

\begin{figure*}
    \centering
    \includegraphics[width=0.65\textwidth, angle=0]{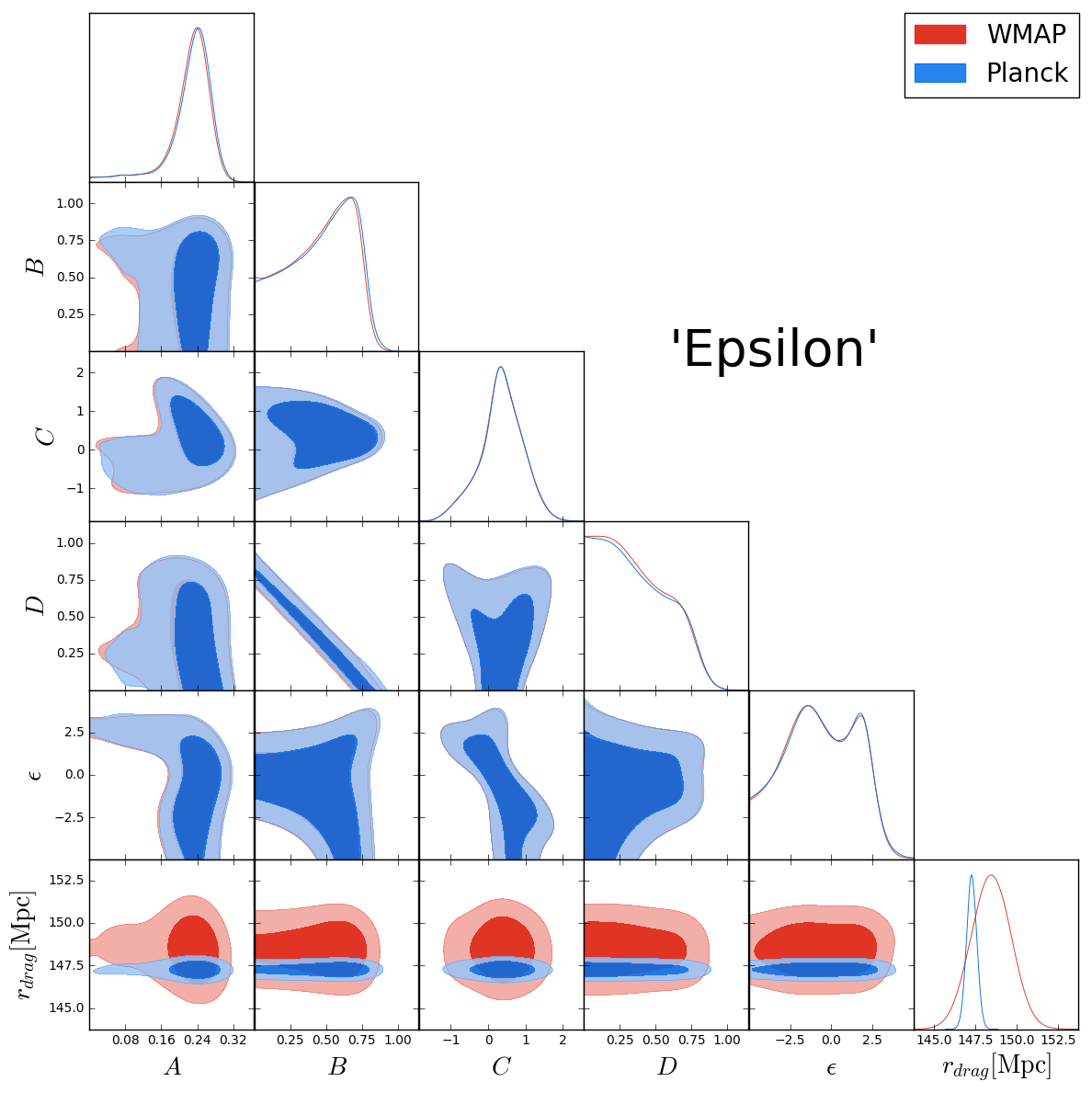} \\
    \includegraphics[width=0.65\textwidth, angle=0]{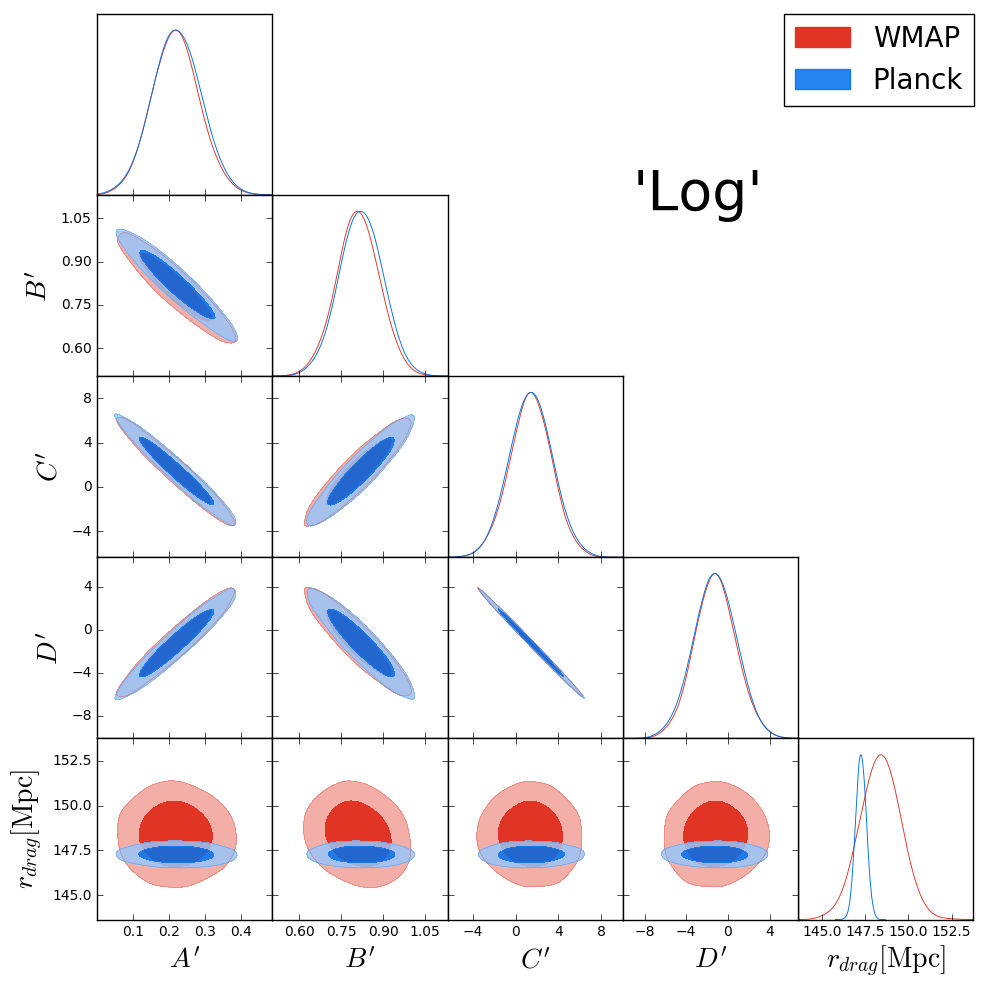}
    \caption{Posterior likelihoods for the `epsilon' (above) and `log' (below) parameterizations of $H(z)$. Blue contours show 68\% and 95\% constraints using the \Planck\ prior on $r_d$. The red contours
(largely hidden by the blue contours) show the constraints using the WMAP prior on  $r_d$.}
    \label{fig:triangle}
\end{figure*}

\begin{figure*}
    \centering
    \includegraphics[width=0.49\textwidth, angle=0]{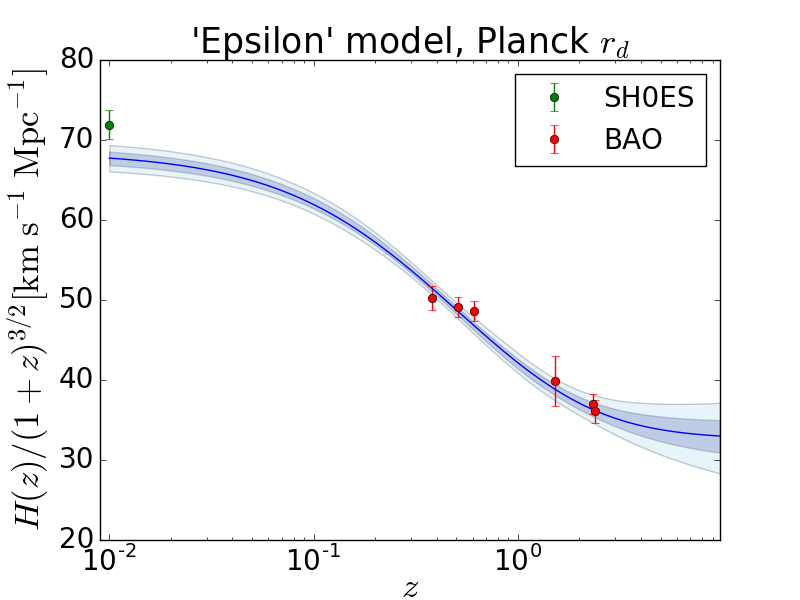}
    \includegraphics[width=0.49\textwidth, angle=0]{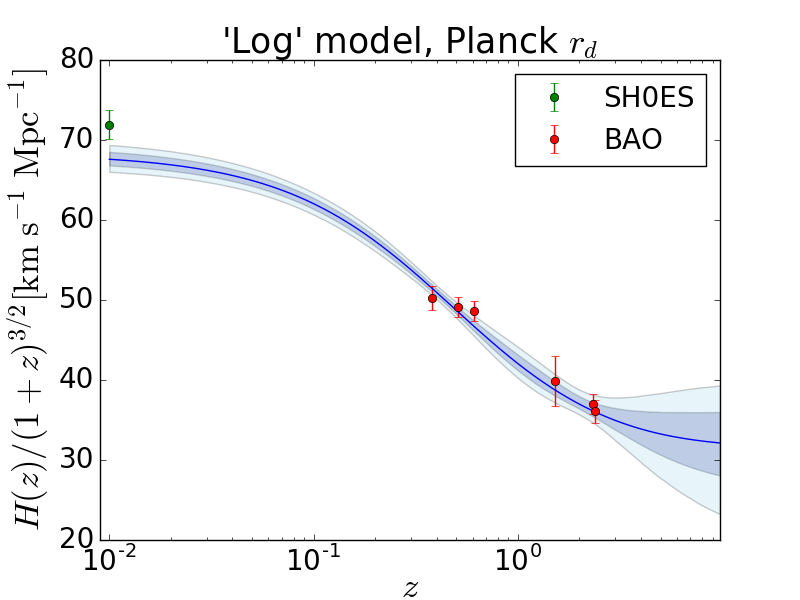} \\
    \includegraphics[width=0.49\textwidth, angle=0]{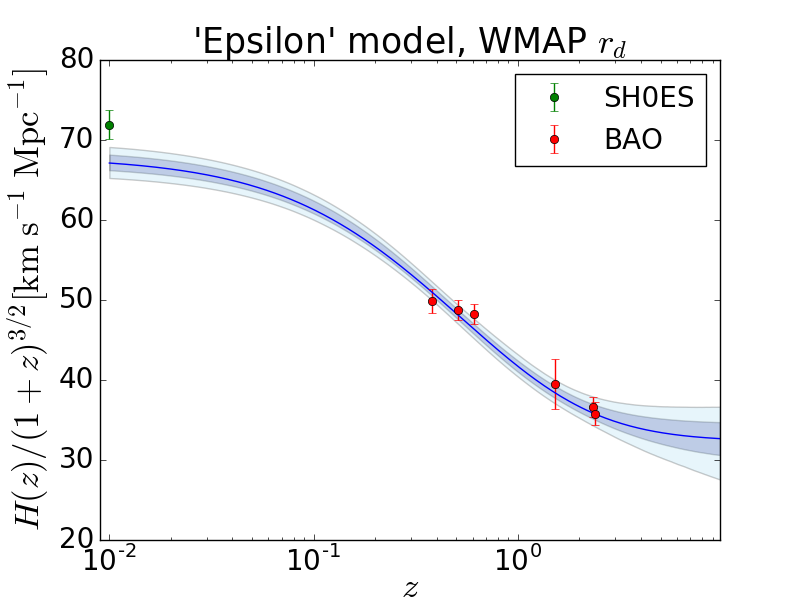}
    \includegraphics[width=0.49\textwidth, angle=0]{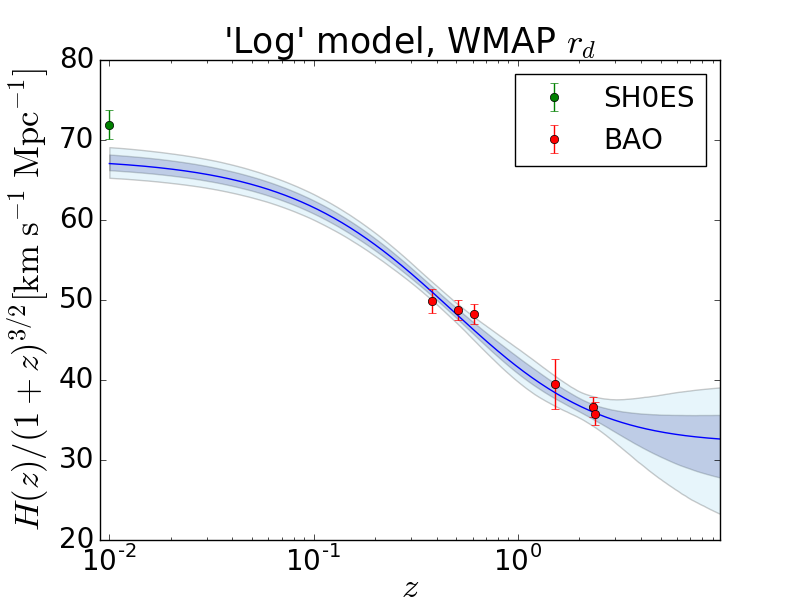} 
    \caption{$H(z)$ reconstruction for the epsilon model (left hand panels) and log model (right hand panels) for the \Planck\ and WMAP priors on $r_d$:  The blue lines show the best fits, and the
 bands show the allowed one and two sigma ranges. The red points show the  BAO estimates on $H(z)$  from Table \ref{table:BAO} plotted assuming the central values of the priors on $r_d$. The 
R18 SH0ES forward distance ladder estimate of  $H_0 = 73.48 \pm 1.66 \Hunit$ is plotted as the green point in each panel.}
    \label{fig:reconstruction}
\end{figure*}

For the supernovae (SNe) data, we use the new Pantheon
sample\footnote{The files are publicly available as a {\tt CosmoMC}
  module at
  \url{http://kicp.uchicago.edu/~dscolnic/Pantheon_Public.tar}.}
\citep{Scolnic:2017}. This dataset contains SNe spanning the redshift
range $0.01 < z < 2.3$ drawn from a number of surveys: The Pan-Starrs1
survey \citep{Rest:2014, Scolnic:2014}, CfA1-CfA4 \citep{Riess:1998,
  Jha:2006, Hicken:2009, Hicken:2012}, CSP \citep{Contreras:2010,
  Folatelli:2010, Stritzinger:2011}, SNLS \citep{Conley:2011,
  Sullivan:2011}, SDSS \citep{Frieman:2008, Kessler:2009}, SCP survey
\citep{Suzuki:2012}, GOODS \citep{Riess:2007} and CANDELS/CLASH survey
\citep{Rodney:2014, Graur:2014, Riess:2018b}. We also used the Joint
Light-Curve Analysis (JLA) sample \citep{Betoule:2014}. The JLA
compilation gives almost identical results for $H_0$ as the Pantheon
sample,  so we do not present those results here.

\section{Results}
\label{sec:results}

\subsection{Constraints on the expansion history}
\label{subsec:Hz}

We use the {\tt CosmoMC}
package\footnote{\url{https:://cosmologist.info/cosmomc}.}
\citep{Lewis:2002, Lewis:2013} to sample the free parameters of the
models. For $r_d$, we adopt Gaussian priors with dispersions as given
in equations (\ref{rd1a}) and (\ref{rd1b}). For the epsilon model, the
parameters $A$, $B$ and $D$ are constrained to be positive and we
impose the constraint that $\epsilon > -5$. For the log model we
impose the conditions that $A^\prime$ and $B^\prime$ should be positive.

The constraints on the parameters of each model are illustrated in Figure
\ref{fig:triangle}. The parameters in the epsilon model show complex
degeneracies in comparison to the parameters of the log model.
Nevertheless, the expansion histories $H(z)$ allowed by the two models
are almost identical as shown in Figure \ref{fig:reconstruction}. The
overall scaling of $H(z)$ is set by the $r_d$ prior. The BAO and
SNe data then strongly constrain the redshift dependence with the SNe 
and are particularly
important in fixing the slope of $H(z)$ at low redshifts (as will be
discussed in more detail below). The epsilon and log models give
almost identical results, differing at redshifts $z> 2.4$ where the
models become unconstrained by the BAO and SNe data.

\begin{figure}
    \centering
    \includegraphics[width=0.5\textwidth, angle=0]{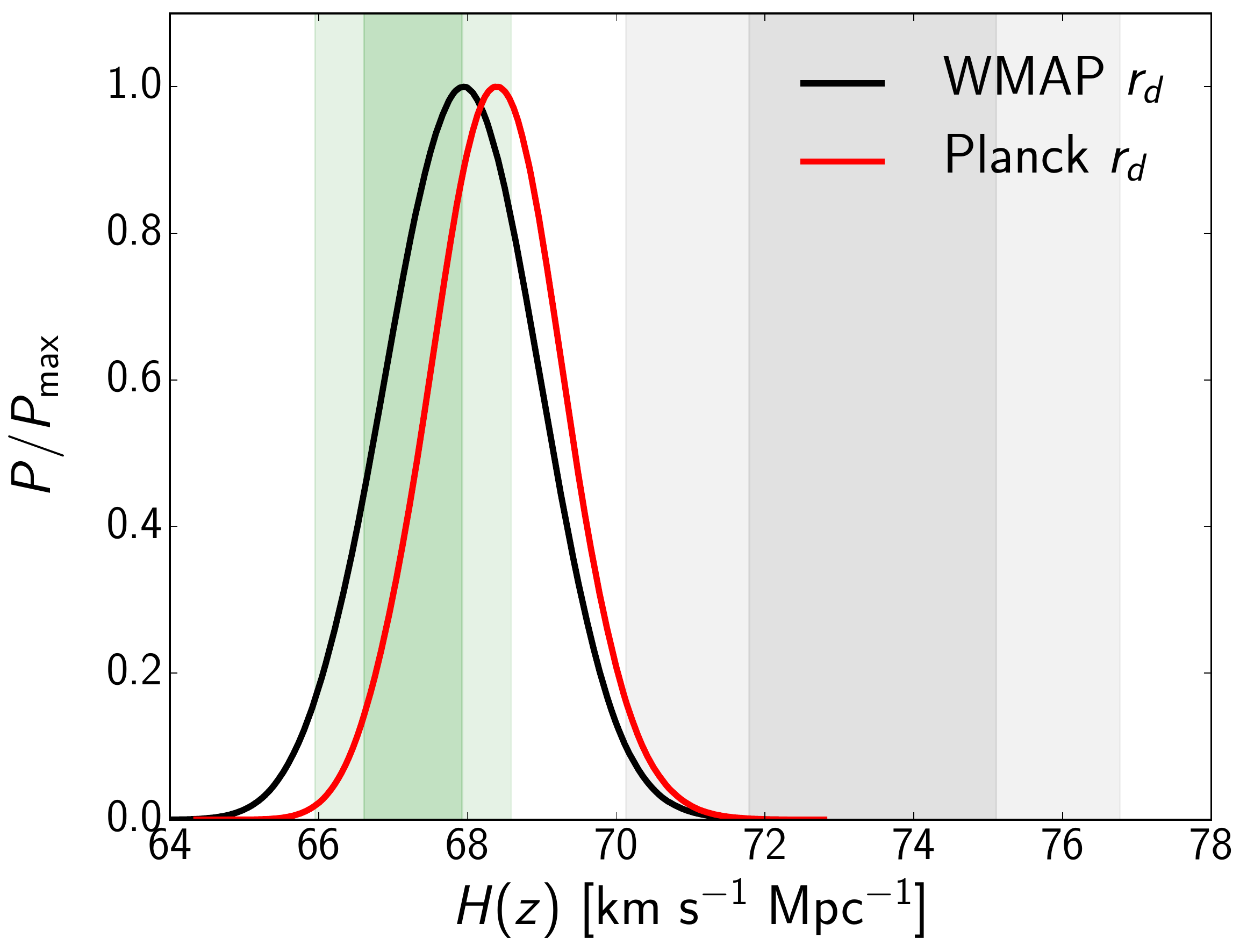}
    \caption{Posteriors for the Hubble constant $H_0$ derived from the epsilon model using 
the WMAP and \Planck\ $r_d$ priors. The grey bands show the one and two sigma errors for the value obtained by R18, while the green bands show the \Planck\ base \LCDM\ value from P16.}
    \label{fig:H0}
\end{figure}

The main results of this paper are illustrated in Fig. \ref{fig:H0} which shows posteriors on $H_0$ for the epsilon model. We find
\beglet
\begin{eqnarray}
H_0 &\hspace{-2mm}=\hspace{-2mm}&  68.42 \pm 0.88 \ \Hunit,   {\rm Planck} \ r_d \ {\rm prior}, \label{H01a}\\
H_0 &\hspace{-2mm}= \hspace{-2mm}& 67.9 \pm 1.0 \ \Hunit,    \ \ \   {\rm WMAP9} \ r_d \ {\rm prior}. \hspace{8mm} \label{H01b}
\end{eqnarray}
\endlet 
The estimate (\ref{H01a}) is about $1\sigma$ lower, and has a smaller error,
than the similar analysis of \cite{Bernal:2016} (which gives
$H_0 = 69.4\pm 1 \Hunit$) because of differences in methodology and
improvements  in the BAO and SNe data.
Both estimates (\ref{H01a}) and (\ref{H01b}) are much closer to the \Planck\ \LCDM\ estimate
of $H_0$ than the SHOES estimate of R18. The value inferred using the
Planck and WMAP $r_d$ priors are respectively $1 \sigma$ and
$0.5\sigma$ higher than \Planck\ estimate and $2.7\sigma$ and
$2.9\sigma$ lower than the R18 value. Evidently, provided General
Relativity is valid, the discrepancy with the R18 estimate of $H_0$ is
unlikely to be a consequence of new physics at redshifts $z \simlt 1$.

 These results are in excellent agreement
with those of \cite{Feeney:2018}, who used an $H(z)$ expansion in terms of the present day values of the
deceleration and jerk parameters,
\begin{equation}
q \equiv - \frac{a \ddot{a}}{\dot{a}^2}, \qquad j \equiv \frac{\dddot{a} a^2}{\dot{a}^3}, 
\end{equation} 
where $a$ is the scale factor of the Friedman-Robinson-Walker metric and dots denote differentiation with respect to time.
Expanding to second order in $z$:
\beglet
\begin{eqnarray}
\left.\begin{aligned}
&H(z)= H_0[ 1 + (1+q_0)z + (j_0 - q_0^2) {z^2 \over 2} ], \qquad \qquad \quad \end{aligned}\right.\label{jerk1} \\
\left.\begin{aligned}
&D_L(z) = {cz \over H_0}[ 1 +  (1+q_0){z\over 2} +  (1- q_0 - 3q_0^2 + j_0) {z^2 \over 6} ].   \label{jerk2}
\end{aligned}\right.
\end{eqnarray} 
\endlet 

\begin{figure}
    \centering
    \includegraphics[width=0.49\textwidth, angle=0]{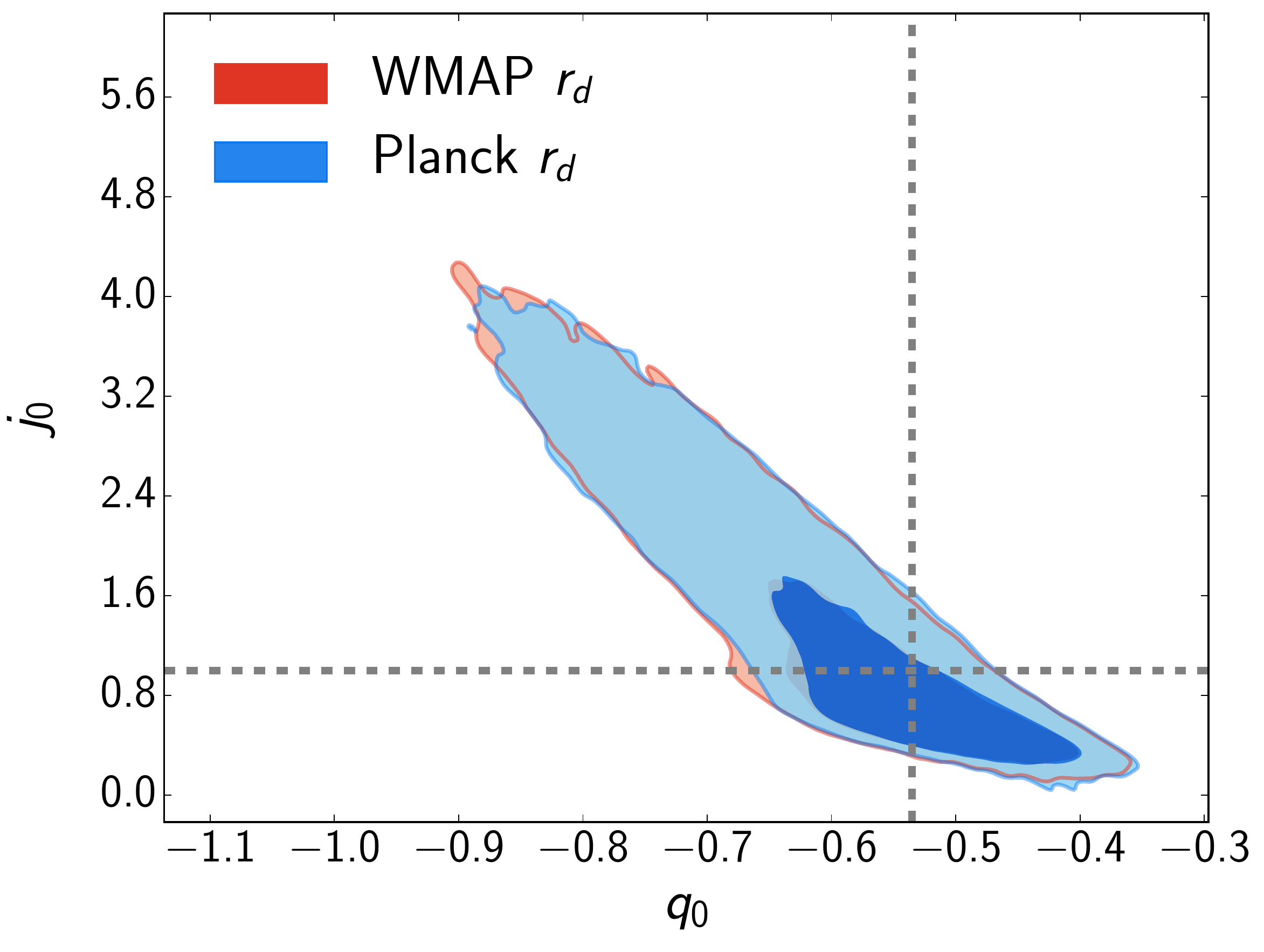}
    \caption{68\% and 95\% constraints on the $q_0$ and $j_0$ parameters determined from the epsilon model. These constraints are set mainly  by the Pantheon SNe sample and
are almost independent of the prior on $r_d$. The lines give the values of $j_0$ and $q_0$ expected in the base \LCDM\ model
with $\Omega_m = 0.31$.  }
    \label{fig:jerk}
\end{figure}

For the base \LCDM\ model with $\Omega_m = 0.31$, $q_0 = 1 - {3
  \Omega_m /2} = -0.535$ and $j_0 = 1$ and the expressions
(\ref{jerk1}) and (\ref{jerk2}) agree well with the exact forms
of $H(z)$ and $D_L(z)$ out to a redshift $z\approx 0.6$ (covering the
redshift range of the BOSS DR12 galaxy measurements). Figure \ref{fig:jerk} shows our
constraints on $q_0$ and $j_0$, which are determined mainly by the
Pantheon SNe sample and so are nearly independent of $r_d$. These
distributions are consistent with the values expected in base
\LCDM. Although these distributions have extended tails, the gradient
$dH(z)/dz$ at low redshifts is tightly constrained by the Pantheon SNe
(Fig. \ref{fig:reconstruction}) which is why it is not possible to
match the BAO $H(z)$ measurements with the SH0ES estimate of $H_0$.
It is also worth noting that the SH0ES methodology matches Cepheid-based distance measurements of SNe host galaxies to more distant
supernovae assuming the relation (\ref{jerk2}) with $q_0 = -0.55$ and
$j_0 = 1$, based on fits to the SNe magnitude-redshift relation. It is
inconsistent, therefore, to apply the R18 $H_0$ measurement as a fixed prior, independent
of the underlying cosmological model, and to infer a cosmology 
that conflicts with the SNe magnitude-redshift relation since the SNe magnitude-redshift
relation is a fundamental part of the $H_0$ determination. This inconsistency
needs to be borne in mind when using the direct measurement of $H_0$ to
set a scale for the sound horizon \citep[e.g.][]{Heavens:2014, Bernal:2016}.

\subsection{Consistency of $r_d$ with high redshift physics}
\label{subsec:rd}

The inverse distance ladder constraints on $H_0$ derived in this paper  assume that there is no new physics at high redshift
that can alter  CMB estimates of $r_d$. BBN provides a strong test of new physics at high redshift and can, in principle,
be used to test the consistency of CMB estimates of $r_d$.  
The most recent estimates \citep{Cooke:2018} of the deuterium to hydrogen ratio $D/H$, based on seven 
low metallicity damped Ly$\alpha$ systems, give
\begin{equation}
{\rm 10^5(D/H) = 2.527 \pm 0.030}. \label{BBN1}
\end{equation}
Assuming three (non-degenerate) neutrino families and BBN, the estimate (\ref{BBN1}) can be converted into a constraint on $\Omega_bh^2$. This conversion is, however, dependent on uncertainties in the $d(p,\gamma)^3{\rm He}$ reaction rate.
\cite{Cooke:2018} use the  theoretical reaction rate from  \cite{Marcucci:2016} and the experimental value
from \cite{Adelberger:2011} to illustrate the sensitivity of $\Omega_bh^2$. They find:
\beglet
\begin{eqnarray}
100 \Omega_b h^2 &=&  2.166 \pm 0.019, \quad {\rm Marcucci \ et \ al.}, \ \ \ \ \ \  \label{BBN2a} \\
100 \Omega_b h^2 &=&  2.235 \pm 0.037, \quad {\rm Adelberger \ et \ al.}, \ \ \ \ \ \ \label{BBN2b}
\end{eqnarray}
\endlet
where the error in (\ref{BBN2b}) is dominated by the error in the \cite{Adelberger:2011} cross-section. The estimate
(\ref{BBN2a}) is lower by $2.4\sigma$ compared to the P16 TT+TE+EE+lowP value of $100 \Omega_bh^2 = 2.225 \pm 0.016$
for the base \LCDM\ cosmology, whereas (\ref{BBN2b}) is consistent with the P16 value to within $0.25\sigma$. We consider
these two values and associated error estimates in the analysis below.

\begin{figure}
    \centering
    \includegraphics[width=0.5\textwidth, angle=0]{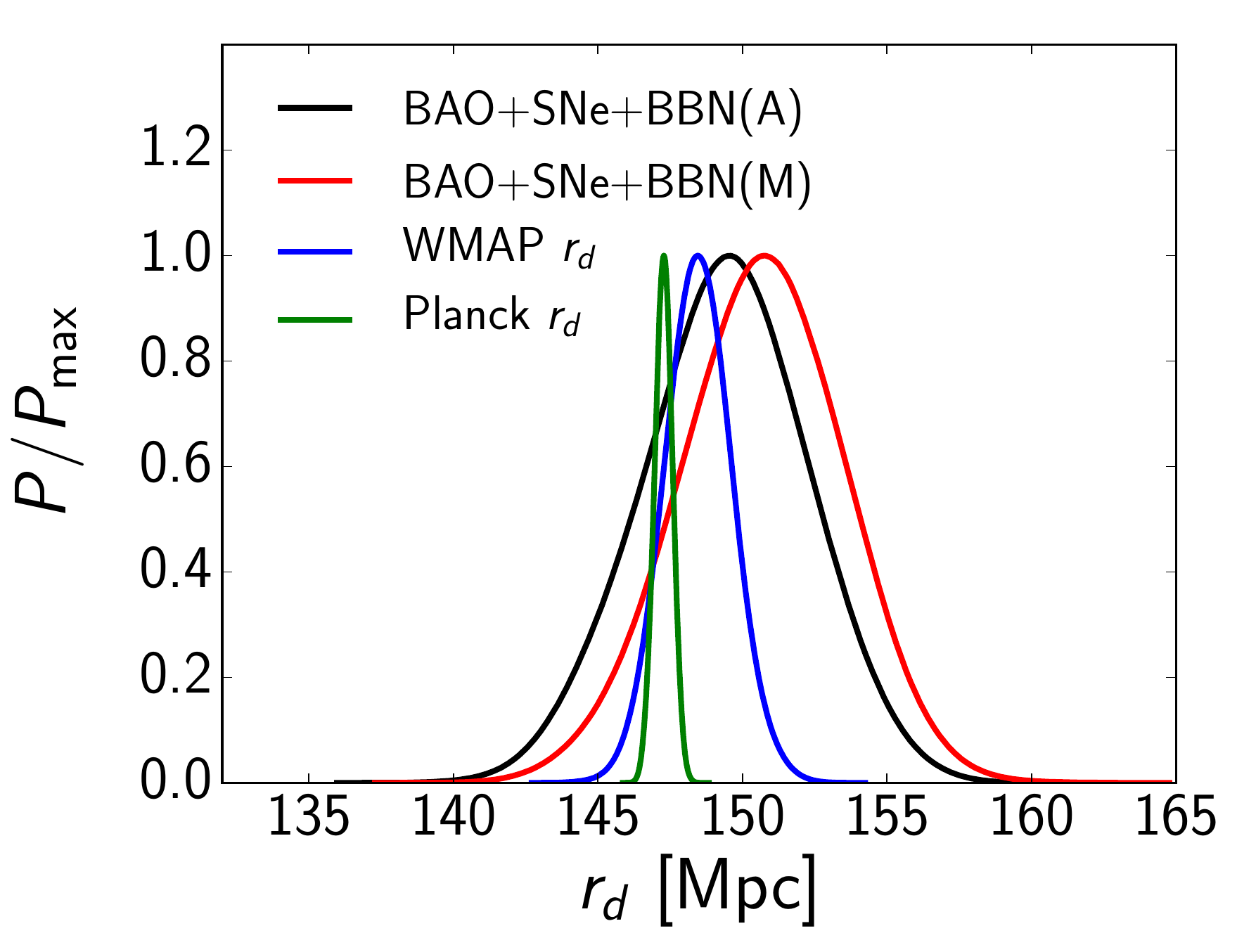}
    \caption{The CMB constraints on the sound horizon $r_d$ from WMAP and \Planck\ used in this paper. The black and red curves
show the posteriors on $r_d$ determined by fitting to the BAO and Pantheon SNe data assuming the base \LCDM\ cosmology and
BBN contraints on $\Omega_b h^2$. The curve labelled BBN(M) assumes the \protect\cite{Marcucci:2016} $d(p,\gamma)^3{\rm He}$
reaction rate (labelled BBN(M)). The curve labelled BBN(A) uses the  experimental rate from  \protect\cite{Adelberger:2011}.}
    \label{fig:rd}
\end{figure}

We then follow \cite{Addison:2018} and \cite{Abbott:2017} in using
these BBN estimates together with supplementary astrophysical data to
infer $r_d$ assuming the base \LCDM\ cosmology. Here we have combined
the BBN constraints with the BAO measurements and the Pantheon SNe
sample, as described in Section \ref{sec:data}. The posteriors on
$r_d$ are shown in Fig. \ref{fig:rd} and are consistent with the $r_d$
constraints from WMAP and \Planck. To the extent that BBN probes early
Universe physics, we find no evidence for any inconsistency with the
values of the sound horizon inferred from CMB measurements.

\cite{Bernal:2016} suggested that the $H_0$ tension can be partially
relieved by invoking extra relativistic degrees of freedom in addition to
the $N_{\rm eff} = 3.046$  expected in the standard model. This solution is
disfavoured by the latest \Planck\ analysis. Allowing $N_{\rm eff}$ to vary
as an extension to the base-\LCDM\ cosmology, \cite{Planck:2018} find $N_{\rm eff} = 2.99 \pm 0.17$, $H_0 = 67.3 \pm 1.1 \Hunit$ and $r_d = 147.9 \pm 1.8\ {\rm Mpc}$ for the TT,TE,EE+lowE+BAO+lensing likelhood combination. Additional relativistic degrees of freedom are therefore tightly constrained by the latest data.

\section{Conclusions}
\label{sec:conc}

The precision and redshift reach of BAO measurements has improved
substantially over the last few years. Together with SNe data, it is
now possible to reconstruct the time evolution of $H(z)$ accurately
without invoking any specific model of the physics of the late time
Universe other than the validity of the FRW metric of General Relativity.  If we assume
that there is no new physics at early times, then CMB measurements
constrain the sound horizon, $r_d$, and this in turn fixes the
absolute scale of $H(z)$ allowing an extrapolation to $z=0$ to infer
$H_0$. Our results disagree with the direct measurement of $H_0$ from
the SH0ES collaboration and are in much closer agreement with the
$H_0$ value determined by \Planck\ assuming the base
\LCDM\ cosmology. This conclusion holds irrespective of whether we use
a prior on $r_d$ from WMAP or from \Planck.

Our results are consistent with previous work on the inverse distance
ladder \citep[e.g][P16]{Percival:2010, Aubourg:2015, Feeney:2018}. In agreement
with  \citep{Bernal:2016}, we reach this conclusion without having to assume any specific
model for the time evolution of dark energy or its interaction with
dark matter and baryons. As long as there is no new physics in the early Universe
that can alter the CMB value of the sound horizon, the new BAO
measurements from BOSS provide accurate absolute measurements of
$H(z)$ in the redshift range $0.38-2.4$. The SNe data then provide a
strong constraint on the gradient of $H(z)$ at lower redshifts, which
is compatible with the gradient expected in the base
\LCDM\ cosmology. The data therefore do not allow a rise in $H(z)$
at low redshift with which to match the SH0ES direct measurement of $H_0$.   We
conclude that it is not possible to reconcile CMB estimates of $H_0$
and the SH0ES direct measurements of $H_0$ by invoking new physics at
low redshifts.

If the tension between the CMB estimates of $H_0$ and direct
measurements is a signature of new physics, then we need to introduce
new physics in the early Universe.  This new physics must lower 
 the sound horizon by about 9\% (i.e. to about
$135\ {\rm Mpc}$) compared to the values used in this paper while preserving the structure of the temperature
and polarization power spectra measured by CMB experiments. This new physics also needs to preserve
the consistency between BBN and observed abundances of light elements. These requirements pose interesting challenges
for theorists.

\section*{Acknowledgements}
Pablo Lemos acknowledges support from an Isaac Newton Studentship
at the University of Cambridge and from the Science and Technologies
Facilities Council. We would like to thank Hiranya Peiris  and 
participants of the Kavli Cambridge Workshop on Consistency of Cosmological Datasets
 for useful discussions. We also thank Licia Verde and Raul Jimenez for comments
on an early draft of this paper.

\bibliographystyle{mnras}
\bibliography{refs} 

\label{lastpage}
\end{document}